# Mechanical Behavior of Axonal Actin, Spectrin, and Their Periodic Structure: A Brief Review

Md Ishak Khan[1], Sheikh Fahad Ferdous[2], and Ashfaq Adnan[1, +]

[1]Department of Mechanical and Aerospace Engineering, University of Texas at Arlington, Arlington, Texas 76019, USA

[2]Department of Applied Engineering and Technology Management, Indiana State University, Terre Haute, Indiana 47809, USA

[+]Corresponding Author. Email: aadnan@uta.edu

**Abstract**

Actin and spectrin are important constituents of axonal cytoskeleton. Periodic actin-spectrin structures are found in dendrites, initial segment of axon, and main axon. Actin-spectrin periodicity has been hypothesized to be manipulating the axon stability and mechanical behavior. Several experimental and computational studies have been performed focusing on the mechanical behavior of actin, spectrin, and actin-spectrin network. However, most of the actin studies focus on typical long F-actin and do not provide quantitative comparison between the mechanical behavior of short and long actin filaments. Also, most of the spectrin studies focus on erythrocytic spectrin and do not shed light on the behavior of structurally different axonal spectrin. Only a few studies have highlighted forced unfolding of axonal spectrin which are relevant to brain injury scenario. A comprehensive, strain rate dependent mechanical study is still absent in the literature. Moreover, the current opinions regarding periodic actin-spectrin network structure in axon are disputed due to conflicting results on actin ring organization – as argued by recent super-resolution microscopy studies. This review summarizes the ongoing limitations in this regard and provides insights on possible approaches to address them. This study will invoke further investigation into relevant high strain rate response of actin, spectrin, and actin-spectrin network – shedding light into brain pathology scenario such as traumatic brain injury (TBI).

# 1 Introduction

There are three (3) major axonal cytoskeletal components in neuron: microtubules (MT) [1] supported by microtubule associated proteins such as MAP1B and tau [2,3], neurofilaments (NF) [4], and microfilaments (MF) [5]. Among these, the structural units of MFs are globular or G-actin, which form filamentous or F-actin. However, these three components do not comprehensively define the axonal structure, as recent seminal super-resolution microscopy study has shown that axonal diameter is determined by periodic arrays of actin ring formed by short F-actin filaments [6]. Furthermore, spectrin, which is structurally and functionally important component of axonal cytoskeleton, forms heterodimers and eventually, tetramers – which generate a lattice by connecting the periodic actin rings [7]. Furthermore, spectrin-membrane association is established by several membrane-associated proteins – such as adducin which caps one end of F-actin and promotes actin-spectrin bond [8,9], ankyrin which binds spectrin to membrane [7], etc.

The overall mechanical behavior of axon and extent of mechanical support to the cytoskeleton are determined by significant contribution of actin-spectrin network. Therefore, mechanical behavior of i) lone F-actin, ii) lone spectrin, and iii) actin-spectrin interaction are required to be determined – which will eventually facilitate the development of a bottom-up, realistic, all-component inclusive computational model of axon. This study attempts to explore the three aspects mentioned above. A thorough literature review is required to comprehend the current understanding and approaches in the field of mechanical behavior of actin and spectrin, and substantiation of the periodicity of actin-spectrin network in the axonal cytoskeleton – which is presented here. In a nutshell, this manuscript presents relevant studies regarding actin, spectrin, and actin-spectrin skeleton.

## 2 Structure of Actin, Spectrin, and Periodic Actin-Spectrin Network

### 2.1 Actin: Structure and Mechanical Behavior

#### 2.1.1 Actin Structure

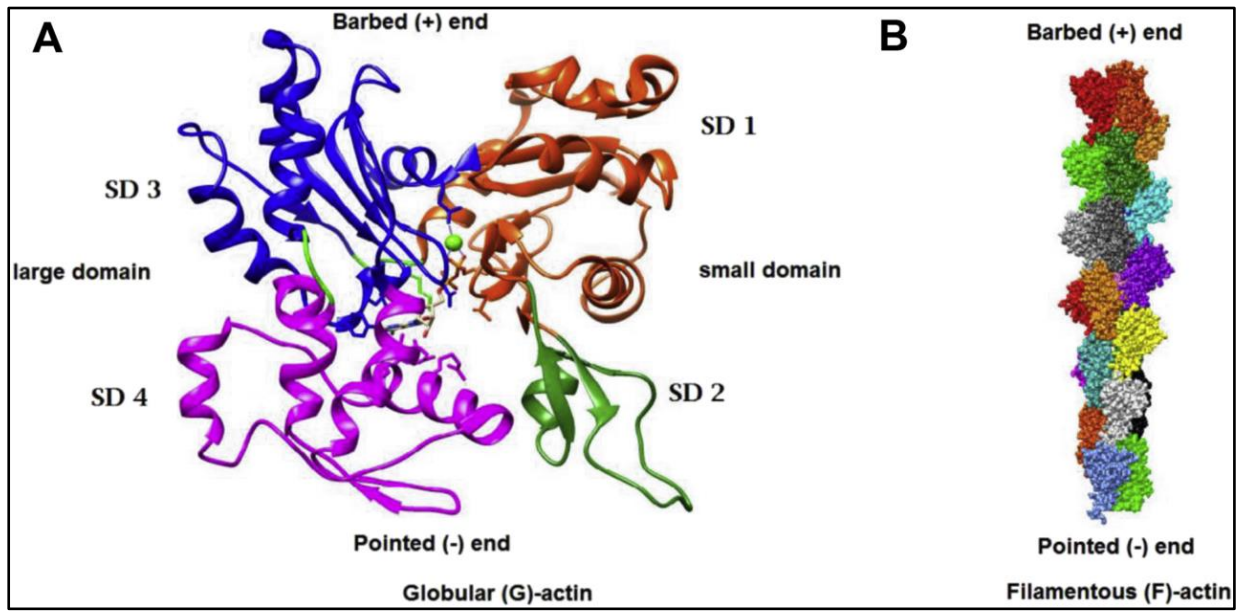

**Figure 1:** Globular or G-actin (left), and filamentous or F-actin (right). The F-actin is believed to contribute to the stability of axonal cytoskeleton. Ref.: [10]

Actin, being one of the most conserved biological structures found in nature has been explored in numerous experimental and computational ways. Although actin has diversified functionality, interaction with cell membrane is one of the most significant attributes of this cytoskeletal component. Actins can be related to several neuropathological disorders. For example, cancer-affected cytoskeleton properties depend on actin filaments to microtubule content [11]. Also, AFM experiment shows that affecting MTs reduces stiffness of axons greatly, while in case of affecting actins the effect is not that severe [5]. However, the later part of this section depicts that due to importance of actin and other constituents of axonal cytoskeleton in regulating overall axonal behavior, numerous experimental and computational studies have been undertaken.

F-actins contain right handed double helices of actin monomer with around $18nm^2$ cross sectional area and several micrometers of length [12]. It has an asymmetric structure with ~6nm long monomer [13,14]. The monomeric subunit (G-actin) has four subdomains, the conformation of which strongly depends on ATP or ADP-bound states [15,16] (Figure 1). X-ray diffraction experiments show that globular to stable, flat, fibrous actin transition can occur by rotation of two major domains [17].

Electron microscopy shows different structural states of actin, especially from the perspective of subunit to subunit connection, and therefore, it should be studied as an ensemble of different structures [18]. Cryo-EM technology has facilitated direct visualization of secondary structure of actin [19]. Immunofluorescence has also been applied for examining actin network structure [20]. Actin networks and bundles should be studied along with the properties of individual filaments, as they show the holistic characteristics of the cytoskeletal component. There are additional dynamic aspects of actin with respect to intermolecular bonding in F-actin, modifiers of the structure, and differences within different monomeric forms of actin – which are thoroughly worked out in recent studies [21]. To maintain focus on mechanical behavior of actin in filament level, discussions on these aspects are not shown in this review.

Furthermore, actin binding proteins play an instrumental role to the assembly and parallel branching network of actins [22]. Also, it is asserted in multiple studies that actin filaments are crosslinked by actin binding proteins (ABP) such as filamin, and a single type of ABP can lead to formation of different actin microstructures [23]. F-actin, being crosslinked by bundling proteins such as actinin, can assemble into frequently branching 3D network [24]. Additionally, divalent cation binding sites induce F-actin polymerization [25]. Therefore, the growth and network formation of actin is highly dynamic – regulated by not only ABPs but also localization of binding sites.

It is worth mentioning that actin network is associated with various cross-linking mechanism which may lead to tightly packed bundles, or loosely connected network. Tightly packed bundles are created by crosslinks of fascin or fimbrin [26,27], while loose networks can be facilitated by crosslinkers like filamin, actinin, or spectrin (for both erythroid and axonal networks) [28,29] or bundles like fimbrin, villin, dematin, etc. [30,31]. However, discussing the mechanical aspects of all of them is out of the scope of the manuscript, and in the later sections, the review will be limited to axonal networks only (short actin interaction with mostly α2-β2 spectrin).

### 2.1.2 Mechanical Properties of Actin

Mechanical properties, especially the persistence length of actin filament have been studied in earlier works [32,33]. The persistence length with all subunits in ATP state (F-ATP) is ~17µm, although it can vary to a large extent. ATP bound, unfolded actin DB-loop persistence length has been found to be twice of ADP bound, folded DB-loop in actin [34]. Further studies revealed that unfolding of subunit structures may lead to alteration of persistence length [17,35]. Evidently, stable state of actin is dependent on the folded DB-loop, which is a result of low free energy. In other words, folded DB-loop leads to formation of softer actin filament, eventually leading to a shorter persistence length [35–37].

Due to having helical structure, bending and twisting of F-actin have significant effects, especially on short ones [38]. However, the structural responses like extension, bending and twisting might be the result of strictly mechanical loading, strictly induced by biochemical parameters, or a combination of both. From that perspective, mechanical behavior of actins become complicated, and it is difficult to decide which aspect to consider for a specific case. For example, in high strain rate loading (TBI scenario), it might be assumed that due to the extreme mechanical loading, biochemical aspects might be overlooked. However, this is not realistic, as in axonal cytoskeletal component, mechanically induced injury may lead to biochemical implications (one example is hyperphosphorylation of tau after TBI) [39]. For actin, there is a specific example in the literature, which suggests that structural changes occur in actin during motor activity, implying the contribution of filament bending flexibility for actimyosin function [40]. Therefore, mechanical loading leading to biochemical implication and biochemical phenomena leading to mechanical property alteration are possible for axonal cytoskeletal components like actin.

One viable approach to address such dilemmas (by characterizing modeling the response to deformation) mentioned in the previous paragraph and study mechanical aspects of such filaments is atomistic simulation e.g. using molecular dynamics (MD) simulation which, however, cannot capture biological phenomena at large length scale. Therefore, an alternative has been coarse-grained (CG) modeling [12,41]. There are several theories regarding behaviors of F-actin, such as elastic rod buckling theory and filament severing by cofilin theory [42,43], while some models have examined mechanics and chemistry interplay [44]. Yogurtcu et al [45], however, has proposed an intermediate scale model by ignoring conformational changes strongly associated with biochemical parameters. For filament length much longer than the helical actin pitch (>1µm), the filament mechanically deforms as a semiflexible rod. Another model worthy to mention considers buckling of actin due to compressive loading [41].

Earlier works have found that bending and torsional rigidities may differ strongly based on ATP or ADP-bound states [32,43,46]. Coarse grain MD simulations validated the findings about the properties like bending and torsional rigidities [47], strengthening the earlier statement in this chapter that biochemical phenomena can affect the mechanical properties of actin [40]. From the force-extension experiment in CG, Chu et al [41] found the stretching stiffness of F-ATP actin and F-ADP actin as 37pn/nm and 31pn/nm, respectively. It is to be mentioned that ATP hydrolysis in actin is not the focus of this study, but the ATP states of actin affects the structure and behavior of actin. For specific aspects ATP hydrolysis, readers may refer to some earlier studies [48,49].

Furthermore, the response to different mechanical loading is also important for actin, as this sheds light on mechanical properties such as stiffness. In relevance, effect of tensile force on actin filaments has been studied earlier [50], and it is shown that tensile force causes decrement of twist angle, leading to increased extensional and torsional stiffness. According to earlier works, torsional rigidity per unit length of actin filament ranges from $2.3 \times 10^{-27}$ Nm$^2$ to $8 \times 10^{-26}$ Nm$^2$ [33,51,52], and extension-torsion coupling is important to comprehend the mechanical behavior of actin. Actin-actin bond breaking force is also measured in some studies, and they showed that turning of filaments greatly reduce the required force (Tsuda et al., 1996). In separate studies, strain dependent behavior of actin networks has been investigated to show that at high strain the elasticity ceases to be linear (Gardel et al., 2004).

Moreover, role of ABPs (which incorporate conformational changes to the structure of actin) on mechanical properties such as stiffness of actin has been investigated by steered molecular dynamics (SMD) simulations [53]. Formation of cofilactin (cofilin bound with each actin monomer) is found to reduce the stiffness of actin filament, especially actin filament with partially bound cofilin (refer to Figure 2). The severing mechanism of cofilin on actin filaments is described in some studies, and found that mechanism of cofilin activity is promoting stress concentrations in junctions of filaments, which is similar to grain boundary fracture of crystalline materials or shear transformation of colloidal materials [54]. Regulatory severing protein (cofilin) has been found to increase bending and twisting compliance of actin in some atomistic and continuum models [37], because buckled cofilactin (actin decorated with cofilin) accelerates severing. Experimentally determined Young's Modulus of actin is reported to be 400MPa-2.5GPa [55–59], to which the MD simulation results of Kim et al [53] and Matsushita et al [50] match closely. Torsional stiffness value obtained from MD simulation performed by Matsushita et al [52] also agreed with the experimental values, assuming that difference can occur depending on the initial conformational state.

Going further into mechanical properties, range of obtained rheological parameters of actin, such as shear modulus and storage modulus from different experiments have been investigated as the reported values differ from 0.01Pa to tens of Pa, and it is concluded that mechanical properties depend on initial length of filament as well as preparation, polymerization condition and storage methodology [60,61]. It is also found in a study we have already mentioned that stiffness of 1µm long actin is higher in case of association with tropomyosin [56].

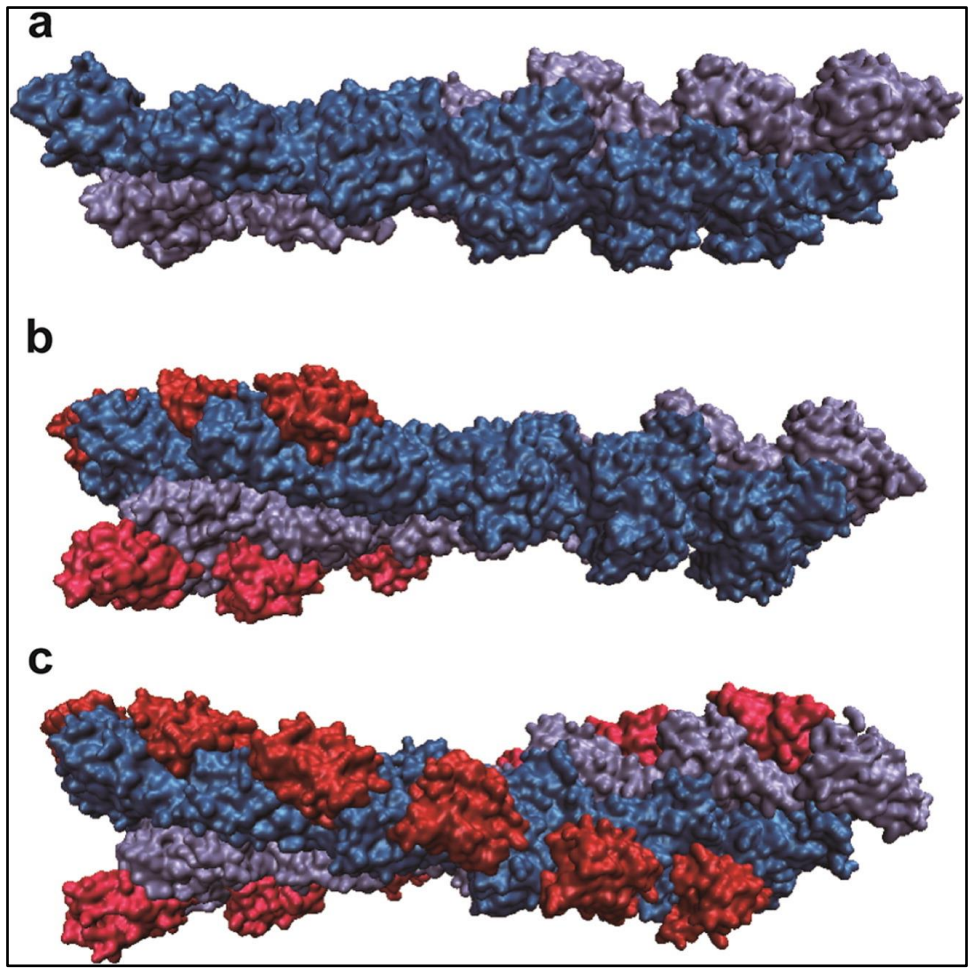

**Figure 2:** Actin filament decoration with cofilin, leading to formation of cofilactin, which facilitated determining enhanced flexibility obtained by cofilactin, investigated by steered molecular dynamics (SMD) approach. The model is prepared from protein data bank (PDB) code 3JOS. a) Actin without cofilin, b) actin decorated by 6 cofilin, c) actin decorated with 12 cofilin. Ref: [53].

In the discussion of mechanical behavior of actin, inclusion of viscoelastic behavior is relevant, especially in network level, which has been studied in different works [23,62] and it is found that the response depends on several parameters, such as i) cross-linker off rate, ii) binding energy, iii) characteristic bond length depending on actin-actin or actin-ABP interaction [23]. The response of the network is a combination of elasticity of the network and force-induced cross-linker unbinding and rebinding [63]. Both oscillatory and shear type experiments showed viscoelastic nature for filamentous and non-filamentous actins [64]. Viscoelastic behavior of actin network has been studied by modeling single filament, cross-linking and incorporating Maxwell properties in FEM, and validated by large strain experiments [65]. The authors would also like to mention that specific protein (such as filamin) induced crosslinking may also lead to significant change to mechanics of actin [66], the detailed discussion of which is out of the scope of this manuscript.

After going through structure, mechanical properties, and mechanical behavior, it is relevant to explore the dynamic aspects and mechanics of actin at axonal cytoskeletal level – as the studies on mechanics and dynamics of actin shed light on their response to deformation and filament-to-filament interaction. Mechanics of F-actin cytoskeleton is dictated by diverse mechanical dynamics of F-actin networks and bundles, and it has been reviewed thoroughly [67]. Specifically, cross-linking dynamics has been strongly related to mechanical properties of actin [68]. Rate of deformation also affects the difference between the stiffness of pure actin and actinin - at higher deformation rate, actinin is found to be stiffer than pure actin [69]. Actin, when associated with cross-linking proteins, act as around 40 times stiffer than pure actin under high deformation rate, although the difference becomes indistinguishable under small deformation rate [69], which is justified by multiple rearranging of cross-link hypothesis.

### 2.1.3 Computational Aspects of Actin Modeling

Computationally, there have been numerous approaches to model and characterize actin. For example, atomistic studies on MFs have used Oda [17] or Holmes [21] model. Due to the nanometer length scale, multiscale approaches have also been proven effective from modeling perspective [70]. The structure, network, and stiffness properties have been determined by multiple studies. However, recently published review study, after performing a comprehensive literature exploration has concluded that the stiffness of microfilaments (including actins) can vary significantly, from a few hundred megapascals to 2.5 gigapascals [71], as actin has been studied under the application of a wide range of strains and strain rates. However, most of the stiffness data come from typical long filament of F-actin, not the short filament present in axonal periodic actin-spectrin network. In order to avoid redundancy in the discussion, the stiffness aspect of actin is limited to Table 1 in the manuscript.

Atomistic and CG studies have facilitated obtaining significant insight regarding mechanical behavior of actin including the effect of actin-severing protein [72], effect of crosslinking [73], and dynamic attributes [12]. Aside from the stiffness and strictly mechanical insights, computational studies have also provided with critical information regarding mechanochemical attributes [74], deformation criteria [45], persistence length [47], torsional mechanism [47], viscoelastic characteristics [65], etc. In this section, however, emphasis will be given to the findings of modeling study mechanical behavior of actin under moderate to extreme strain rate which are relevant to brain injury scenario.

From the modeling perspective, there have been numerous MD, coarse grained MD, FEM and continuum scale models on actin, and it is relevant to mention the aspects of some studies we have already cited in the earlier seciton. Fully atomistic models by fitting known structures of G actin has been proposed earlier [17,19,25]. Some MD simulation studies have attempted to capture global structure as well as internal stereochemistry of actin [25]. Coarse-graining from fully atomistic simulations further revealed holistic properties of F-actin [12]. Actin network properties has also been studied by continuum models, which highly emphasize cross-linking proteins [65,75]. Oda model and Holmes model have been used extensively by others in MD simulations to further investigate F actin network structure and properties [35]. Last but not the least, a very recent MD simulation study has found that when subjected to extreme strain rate, actin filaments can show high tensile stiffness [76]. Also, such actin filaments behave as stiffer material with the increase of the applied strain rate. Therefore, computational studies provide evidence that the most important parameters to dictate mechanical behavior of actin are the filament length and applied strain rate.

In this discussion on actin, it is clear that the literature is highly enriched on the mechanical behavior of actin, but most of the studies are relevant to typical long actin filaments. However, as it will be discussed in the later section on actin-spectrin periodic network, the current limitation will be conspicuous – that there are only a few studies that focus on the mechanical behavior of short actin filaments contributing to the actin-spectrin lattice, and quantitative comparison between the contribution of longer versus shorter actin filaments in axon stability is currently absent in the literature. Therefore, further studies on this topic will have high impact in the injury biomechanics and TBI research areas.

## 2.2 Spectrin: Structure and Mechanical Insight

### 2.2.1 Structure of Spectrin

As a significant axonal cytoskeletal component and a member of F-actin crosslinking superfamily, the functionality, structure, and attributes of spectrin have been explored in detail [7]. Structurally, spectrin forms α-spectrin and β-spectrin heterodimers [77] which lead to tetramers [78], eventually forming a hexagonal lattice when combined with periodic actin rings. Among the two α and five β isoforms, the α-II (genetic encoding: SPTAN1) and β-II (genetic encoding: SPTBN1) are the most relevant ones for axonal spectrins [79]. While the discussion on contribution of the individual α-II and β-II in axonal stability is not in the scope of this review, it is interesting to note that the comparative importance of them has been analyzed, and the loss of α-II spectrin leads to more axonal degeneration than that of β-II [79].

It is worth mentioning that due to being part of heterodimer, α and β shares only 30% similarity in structure. Furthermore, they have distinct function in axonal pathfinding [80]. The role of spectrin is recognized in maintaining axon stability and mechanical properties [81,82]. Lack of spectrin has been marked as a source of axon instability, even breaking [83]. As the literature is enriched with numerous studies on erythroid spectrin and the current study focuses on the limitation on axonal spectrin behavior, the term "spectrin" will mean "axonal spectrin" onwards in this manuscript unless otherwise stated specifically (like erythroid spectrin).

Recently proposed medium resolution zipper model of spectrin dimerization shows that α20-21 and β1-2 repeats create a dimer initation site and close the dimer by utilizing electrostatic interaction. At the junction between α20-21 and β1-2 repeats, there are the actin binding domain, adducin binding spot, and $Ca^{2+}$ binding EF hand. The determination of building blocks of spectrin i.e. α and β subunits have been performed experimentally in the early 80's, mostly by using gel-filtration and ion-exchange chromatography, which laid out the groundwork of membrane associated actin-spectrin cytoskeleton [84,85]. However, the function of spectrin was observed from biochemical perspective by incorporating phosphorylation and resulting change in association or dissociation from membrane which leads to specific stability states of the cytoskeleton, instead of response to mechanical loading perspective. It is to be mentioned that the actin-binding domain in spectrin subunits has also been determined in early chromatography researches [86,87] and later green fluorescence microscopy [88], which led to quantification of specific domains and repeat regions later, even for non-erythroid spectrins [89,90]. Additionally, the ~180nm periodicity of spectrin, which is consistent with the periodicity of axon rings placed ~180nm interval along the length of axon has been established by recent nanoscopy studies – which provide concluding evidence that indeed axonal cytoskeleton contains periodic lattice of actin-spectrin network [91], while the exact details of the structure will be dependent on appropriate imaging method.

### 2.2.2 Limited Mechanical and Modeling Insight into Spectrin

Most of the published work on spectrin properties investigate erythroid spectrin and actin-spectrin biochemical interaction [92–97], large deformation and elastic response of erythroid spectrin [98], network level elasticity in erythrocytes [99], etc. These studies provided excellent validation of continuum models of red blood cells by providing reliable length scale relationships. The extent of progress in erythrocyte related modeling which incorporates properties of spectrin are diverse and advanced in literature due to advancements in optical tweezer experiment methodologies and atomic force microscopy (AFM), while the literature lacks insight regarding axonal cytoskeletal modeling of spectrin [100,101].

However, due to biochemical phenomena or mechanical loading, axonal spectrin may unfold and stretch, which leads to the failure of the filament. Essentially, such failures are relevant to traumatic brain injury scenario (TBI). As a biological material, the failure behavior of spectrin can be explained from different viewpoints, such as mechanical, biochemical, or a combination of both [102]. However, forced unfolding research studies are particularly relevant to the current work, as they provide insight regarding extensibility of multi-domain proteins such as spectrin (refer to Figure 3). One example is forced unfolding of tandem spectrin repeats at low forces performed by AFM, which has suggested that tandem unfolding differs significantly from single unfolding [103]. It is to be mentioned that similar insight regarding axonal spectrin is not available in the literature, and it is expected that the mechanical behavior will be significantly different due to the difference in persistence length, stiffness, and packing mechanism of tandem spectrin and axonal spectrin. However, for modeling purpose of the injury scenario of axonal spectrin, one of the focal points will be the unfolding mechanism of spectrin under moderate to high strain rate. This implicit insight will hopefully invoke further molecular level studies on axonal spectrin and actin-spectrin periodic lattice. Only then, a quantitative comparison between the mechanical behavior of axonal spectrin and erythrocytic spectrin will be possible.

Investigating more into mechanical response of spectrin, single molecule force spectroscopy (SMFS) has demonstrated single unfolding of spectrin specifically, which showed that single spectrin unfolding occurs in a stepwise fashion when susceptible to stretching, substantiating the presence of multiple intermediate repeat region in the structure [104]. Also, AFM study has quantified that force required to unfold spectrin repeats ranges between 25-35pN [105]. In addition, biochemical analyses of spectrin folding and unfolding mechanism have attributed different folding mechanism to interchain binding aspects [106], different kinetic characteristics [107], and existence of critical extension [108].

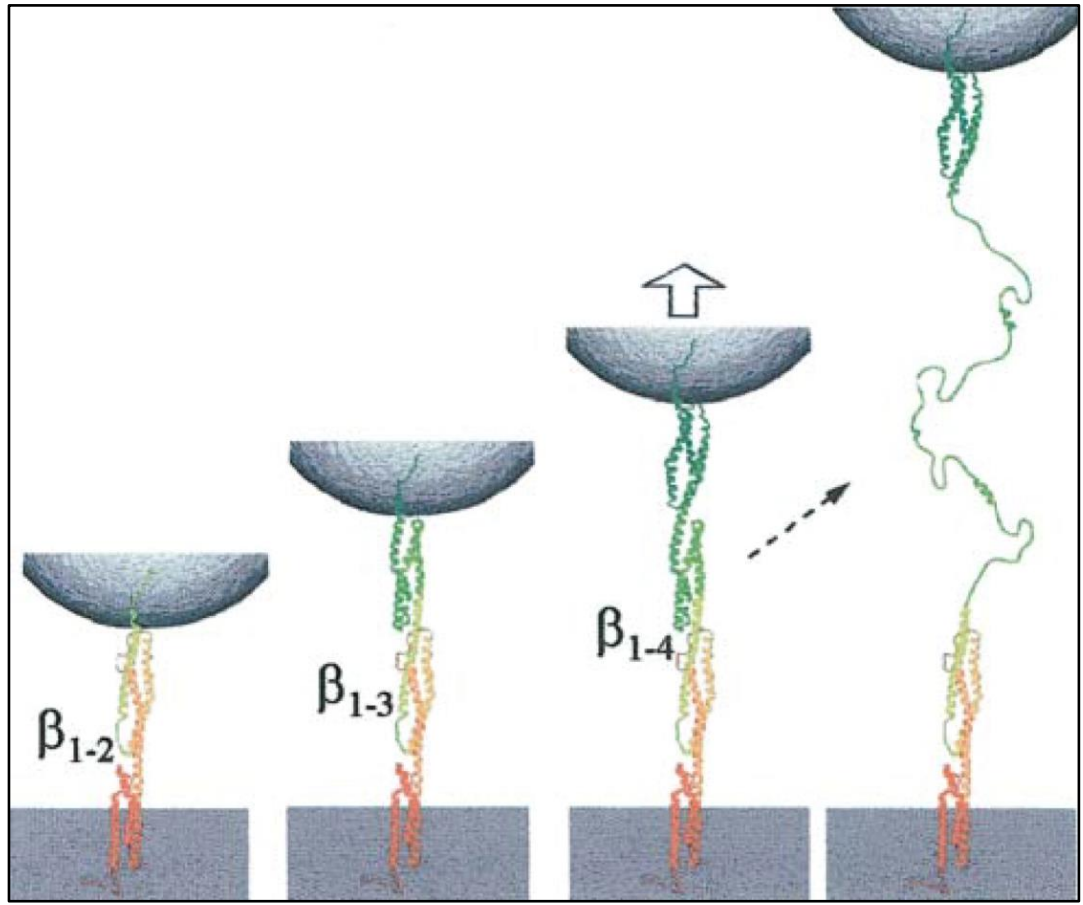

**Figure 3:** Depiction of forced unfolding of spectrin tandem repeats as found by atomic force microscopy (AFM). Ref: [103].

Additionally, steered molecular dynamics (SMD) studies on chicken brain spectrin have examined forced unfolding of multiple repeat spectrin suggesting that α-helical linker can be ruptured if susceptible to forced unfolding, the propagation of which may lead to destabilization of the tertiary structure [109]. Moreover, the same group has investigated the rupture criteria of spectrin specifically by implementing non-equilibrium MD simulation, which has shown that force-extension response changes significantly at ~0.4nm extension, and spectrin behavior can make a transition from elastic to viscous material as suggested by force-extension curves [110]. Formation of stable intermediate unfolded structures has been substantiated by similar MD simulation works on spectrin [111,112]. In this regard, a latest contribution is an MD simulation study that applies high strain rate on α-spectrin and β-spectrin, both of which are found to have high stretchability, which is manifested at high strain rate [76].

Furthermore, theoretical modeling of spectrin network in axonal cytoskeleton has led to insights regarding the extension and fluctuation characteristics, which suggested that spectrins, within the periodicity of axons, can fluctuate in-sync. This study substantiated that a spectrin network can be considered as a slender structure which can be coarse-grained [113].

Lastly, among other mechanical properties of spectrin, the frequency dependence of shear response has been measured by multiple lumped resonating viscoelastometer, which suggested that spectrin dimer can extend at a specific ionic strength [114]. Moreover, viscoelastic and mechanochemical characteristics of spectrin gel have also been determined, but only for erythrocytic ones, and therefore, insight regarding axonal spectrin is still lacking to date [115,116]. The limitation in the literature is therefore evident that the stiffness and mechanical response data are almost limited to erythrocyte related studies [117].

**2.3 Periodic Actin-Spectrin Skeleton: Role in Axon and Strain Rate Dependent Scenario**

The periodic actin-spectrin cytoskeleton structure (Figure 4 and Figure 5) as well as the difference between cytoskeleton structure in dendrites, synapse, axon initial segment, and axonal cytoskeleton have been substantiated only after 2010s by dint of super-resolution microscopy and fluorescence nanoscopy [6,118–122]. The significant advancement in the microscopy front is understood by recent review work and remodeling study on periodic actin-spectrin network [123,124]. Specifically for axonal cytoskeleton, the periodicity is stated to be formed at the early stages of development and extends from proximal to distal end of axon [125]. It is to be mentioned that the periodicity of cytoskeleton is found in different types of cells and across species [126] and even axonal actin structure may differ from rings to waves and trails [127] – but in this section, the discussion is limited to axonal cytoskeleton. For clarity, it is to be mentioned that the actin-spectrin network is a part of the actimyosin network, and therefore, myosin may have important role in determining axonal behavior [128–131]. Nevertheless, as only axonal damage is the focus of this review, the effect at myosin level is not included here. Rather, the contribution of the main individual elements are discussed here, as in preliminary modeling, only the connections between the main constituents such as actin and spectrin will likely to be considered.

Specifically, the periodic actin-spectrin network is found in relevant neuronal structures such as main longitudinal portion of axon, initial segment of axon, and necks of dendritic spines – which differs than the structure of actin found in dendrites changing the conformation as one moves forward along the length of dendrites [132,133]. In this regard, earlier experimentations have investigated spectrin-actin gel elasticity as a function of protein concentration, which proposed actin fiber network crosslinked by spectrin networks at regular intervals [134]. The role of actin-spectrin network in maintaining axon diameter and MT stability have also been well-established by multiple studies including recent ones [8,135–137]. Where it might be intuitive that as the periodic actin-spectrin lattice is constituted of short filaments of actin and longer filaments of spectrin, the mechanical response should be dictated by spectrin. However, the real axonal response is a result of highly dynamic network with involvement of and contribution of multiple membrane-associated proteins and actin-capping proteins aside from actin, spectrin, and their interaction. Therefore, it is relevant to study the recent standings of the actin-spectrin interaction studies. However, the current understanding of axonal actin-spectrin network is enigmatic, as suggested by conflicting results from recent super-resolution microscopy studies [138]. While one study suggests that actin-spectrin network is consisted of short filaments of actin capped by adducin [6], the opposing study asserts that adducin not only acts as an actin-capping protein, but it may attach itself to the side of actin to promote longitudinal interaction. Also, it suggests that rather than a bunch of short filaments of actin, the periodic ring consists of intertwined (braided) long filaments of actin [139]. As the mechanical behavior of actin depends on the length of the filaments, it can be inferred that the mechanical behavior and response to injury of actin-spectrin network is not conclusively defined yet. Therefore, axonal actin-spectrin network remains an interesting topic for further study. Also, it is worth mentioning that not only the actin-spectrin network provides mechanical support to axonal cytoskeleton, but also, they have functionalities related to signaling and axonal transport. Many of such functionalities are activated by receptors as depicted in a recent super-resolution microscpy study on RTK activation in neurons [140]. Such study suggests that membrane-associated cytoskeleton facilitates a dynamic framework of signaling pathways, and thus further investigation in this regard might generate interesting insights.

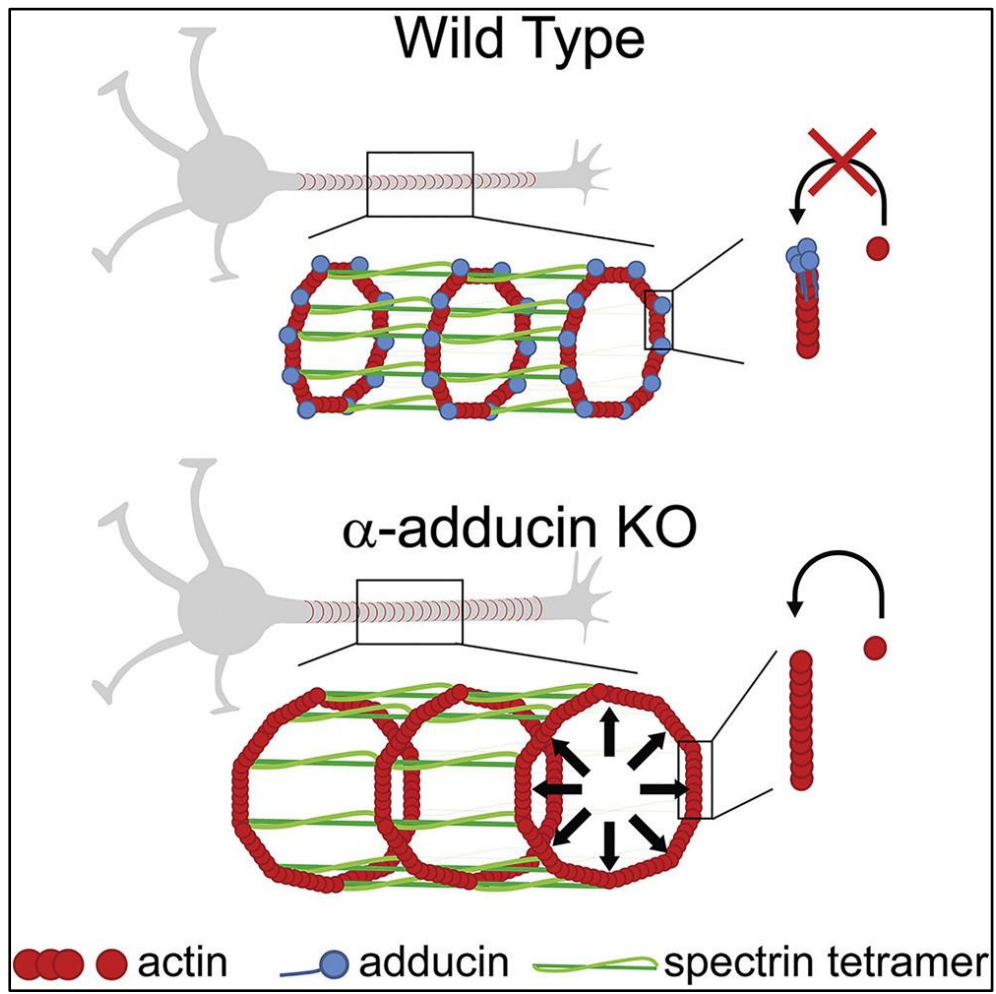

**Figure 4:** Representation of periodic actin-spectrin network, which provides stability to axonal cytoskeleton in cooperation with other proteins such as adducin. Ref: [8]. The purpose of this figure is to represent the periodicity of actin-spectrin lattice in axon, but it also highlights the importance of actin-binding protein such as adducin, the scarcity of which leads to aggressive axon degeneration and enlargement as depicted. The other membrane associated proteins such ankyrin are not shown here.

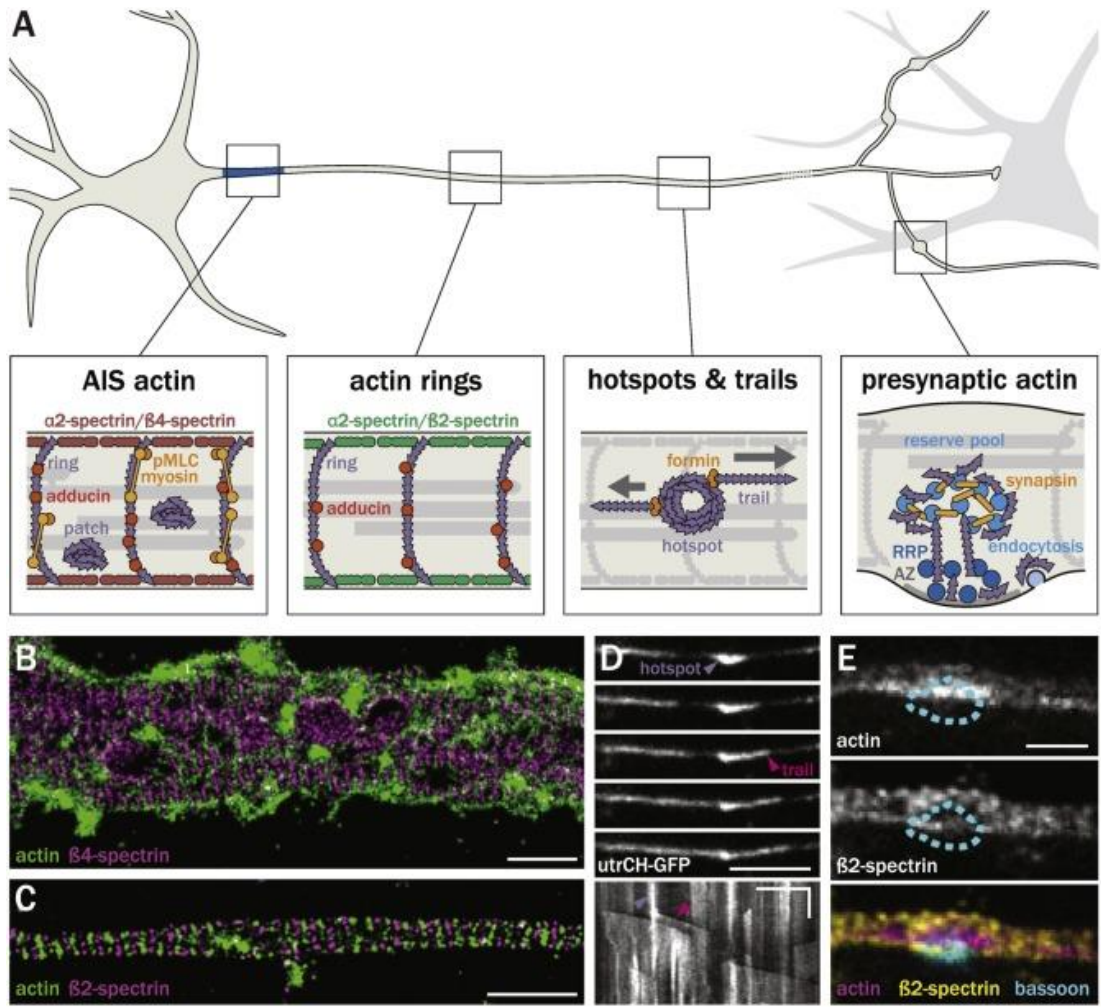

**Figure 5:** Depiction of axonal schematic and results of recent super-resolution microscopy which shows that the actin-spectrin network differs along the longitude of the axon [138]. In panels B (axon initial segment containing α2-β4 tetramer of spectrin) and C (axon distal segment containing α2-β2 tetramer of spectrin), actin and spectrin are stained green and magenta respectively, which clearly suggests the periodicity of actin-spectrin network and their non-uniformity along the length of axon. Panel D shows a live imaging sequence, while E shows actin and spectrin specifically.

Spectrin-actin interaction and associated alteration of cytoskeleton (even axon degeneration) have been found to be manipulated by numerous biochemical agents and phenomena, such as effect of dematin [141], tropic deprivation (TD) [142], etc. Furthermore, continuous remodeling of neural cytoskeleton has been attributed to neural growth, which is dictated by actin-MT and actin-spectrin interaction [143]. However, studies focusing on spectrin-actin interaction or alteration of the interface between them due to mechanical loading is scarce in literature. Among the few recent studies relevant to actin-spectrin cytoskeleton, one has attributed the tension-buffering shock absorber mechanism of the cytoskeleton (especially reversible unfolding of the repeat domains of the spectrin tetramers) to the ability of axon to stretch significantly by performing stretching experiment on chicken dorsal root ganglion in a customized force apparatus [144]. In relevance, red blood cell (RBC) coarse grain modeling has revealed that spectrin contributes to shear stress at lower shear strain, but lipid membrane also contributes at higher strain rates [145].

The knowledge gap in the mechanical behavior of axonal actin-spectrin network is also admitted in another study which combines atomic force microscopy (AFM) and molecular dynamics (MD) simulation approaches as an attempt to distinguish among the somatic, dendritic, and axonal stiffness [146]. This study demonstrates that the axonal stiffness is significantly higher than those of the somatic and dendritic regions. The associated MD simulation shows that coarse graining of the cytoskeletal components can reliably reproduce experimental finding in this regard, and therefore, it can be one of the plausible approaches to address the ongoing limitations. However, the authors also admit that there is significant difference between red blood cell (RBC) associated and axonal actin-spectrin network, and no quantitative comparison between their mechanical properties or behavior does not exist till date. In this regard, a very recent MD simulation study attempts to shed light on axonal actin-spectrin interaciton subjected to high strain rate, and finds that the mode of failure (separation of spectrin from actin) differs based on the applied strain rate: at lower strain rate, the actin-spectrin interface is susceptible to failure, while at higher strain rate, the likely scenario is failure of spectrin filament due to significant stretch [76].

In short, current literature lacks insight in mechanical behavior of axonal actin, spectrin, and actin-spectrin interaction. However, recent studies on axonal cytoskeletal components have focused on applicable strain rate on soft biomaterials [147–149] relevant to brain and different cytoskeletal components such as microtubules, tau proteins, and neurofilaments [150–153]. Such extreme high strain rate scenarios can be captured by undertaking atomistic computational approaches which can be extended to other axonal cytoskeletal components to provide novel insights regarding the specific mechanical behavior of them at extreme strain rate. However, to accomplish these objectives, different customized approaches are expected, such as atomistic based continuum modeling, coarse-graining, adopting comprehensive multiscale maneuver [154], etc. Optimistically, the possible future directions mentioned above will play an instrumental role in developing a bottom-up axon model focusing on moderate to high strain rate scenario and contribute to the existent computational axon models.

## 3  Conclusion

In this study, current scenario and insights regarding actin, spectrin, and actin-spectrin combination are briefed. As biological materials, relevant literature from biochemical perspective is also represented. In this way, it can be asserted that several ongoing limitations from mechanical perspective have been pointed out, which can be summarized as:

Actin: Mechanical insights are mostly present for axonal actin. However, comparative mechanical behavior of short versus long actin filaments are not present in the literature.

Spectrin: Little mechanical insight is present, but mostly for erythroid spectrin, not axonal ones. Erythroid spectrin structure significantly differs from the axonal ones, and therefore, the existent literature does not shed light on mechanical behavior such as forced unfolding maneuver in axonal spectrin.

Actin-Spectrin Network: Little mechanical insight is present, although response to high strain rate is relevant to brain injury scenario. The organization of actin ring is disputed – short adducin-capped actin filaments versus long braided actin filaments in the ring.

Especially, the existent literature fails to cover the high strain rate response of actin-spectrin network. However, the limitations can certainly be attributed to lack of advanced imaging techniques and customized mechanical experimentation as well as modeling. Over the next few years, it can be hoped that the current limitations will be overcome to a significant extent – as structural insights will be more substantiated and microscopy methodologies will be advanced further. Due to the nanometer length scale, MD simulation and associated computational approaches will also serve valuable complementary purpose.

In this study, we have identified some prominent knowledge gaps regarding actin, spectrin, and periodic axonal actin-spectrin network which require immediate attention. Clearer insight regarding such structures and their mechanical behavior awaits further improvement (or additional studies) of super-resolution microscopy. In the meantime, the computational (modeling) approaches will be able to provide further insight regarding their behavior. This study has given emphasis particularly on computational modeling as recent studies has strengthened the view that atomistic simulation is a viable approach to obtain insight regarding axonal cytoskeletal components in traumatic brain injury (TBI) scenario. The culmination of such studies is a comprehensive bottom-up modeling of axon which provides detailed insight of axonal level response to deformation when susceptible to injury. Therefore, this study not only points out the state of the art of the field, but also suggests viable approach for improvement.

Table 1 summarizes the mechanical properties, behavior, or insight available in the literature – substantiating that there is sufficient literature focusing on actin behavior, but not for spectrin and actin-spectrin periodic lattice.

**Table 1:** Actin, Spectrin, and Actin-Spectrin Network: Mechanical Properties and Behavior Found in Literature

| Axonal Cytoskeletal Component | Mechanical Property or Attribute Found | Reference |
|---|---|---|
| Actin (Mostly for typical long F-actin. Quantitative | Persistence length | [32–34] |
| | Unfolding and stretching characteristics | [17,35] |
| | Mechanics of F-actin bundles | [67] |
| comparison between short and long actin filaments is currently unavailable in the literature) | Bending and torsional rigidity and stiffness, twisting | [32,38,46,47,52] |
| | Severing mechanism and response to tensile loading | [37,41–43,50,53,54,72] |
| | Mechanics + chemistry aspects | [44] |
| | Young's modulus | [50,53,55–58,71,155] |
| | Viscoelastic behavior | [23,62–65] |
| | Cross-linking dynamics | [68,69] |
| | Rheological parameters (such as storage and lost modulus) at filament and network level | [60,61,156] |
| | Actin-actin bond | [33] |
| | Strain-rate dependent behavior | [157] |
| Spectrin (mostly for erythroid spectrin. Only the "Forced unfolding mechanism" contains | Shear response | [93] |
| | Elastic response | [98,99] |
| | Biochemical failure | [102] |
| | Forced unfolding mechanism | [103–109,111,112] |
| literature on axonal spectrin) | | |
| Actin-Spectrin Network | Actin-Spectrin lattice organization (conflicting results regarding actin ring) | [6,139] |
| | Contribution of reversible unfolding of spectrin at low strain rate | [144,145] |

## Data Availability

No customized code or program was used in this review study.

## Acknowledgements

This work has been funded by the Computational Cellular Biology of Blast (C2B2) program through the Office of Naval Research (ONR) (Award # N00014-18-1-2082- Dr. Timothy Bentley, Program Manager).

## Author Contributions